\documentclass[useAMS,usenatbib]{mn2e}

\usepackage{graphicx}

\title[New mechanism of radiation polarization in Seyfert-1 AGNs]{New mechanism of radiation polarization in Seyfert-1 AGNs}
\author[N.A. Silant'ev, Yu.N. Gnedin,  M.Yu. Piotrovich, T.M. Natsvlishvili, S.D. Buliga]{N.A. Silant'ev \thanks {E-mail: nsilant@bk.ru}, Yu.N. Gnedin \thanks {E-mail:gnedin@gao.spb.ru}, M.Yu. Piotrovich, T.M. Natsvlishvili, S.D. Buliga \\
Central Astronomical Observatory at Pulkovo, 196140, Saint-Petersburg, Russia}

\begin{document}

\date{Accepted for publication in MNRAS.}

\pagerange{\pageref{firstpage}--\pageref{lastpage}} \pubyear{2016}

\maketitle

\label{firstpage}

\begin{abstract}
In most  of Seyfert-1 active galactic nucei (AGN) the optical linear continuum  polarization degree is usually small (less than $1\%$) and the polarization position angle is nearly parallel to the AGN radio-axis. However,  there are many types-1 AGNs with unexplained intermediate values for both positional angles and polarization degrees. Our explanation of polarization degree and positional angle of Seyfert-1 AGNs focuses on the reflection of non-polarized radiation from sub-parsec jets  in optically thick accretion discs. The presence of a magnetic field surrounding the scattering media will induce Faraday rotation of the polarization plane  that may explain the intermediate  values of positional angles if there is a magnetic field component normal to the accretion disc. The Faraday rotation  depolarization effect in disc diminishes the competition between polarization of the reflected radiation  with the parallel component of polarization and the perpendicular polarization from internal radiation of disc (the Milne problem) in favor of polarization of reflected radiation. This effect allows us to explain the observed polarization of Seyfert-1 AGN radiation even though the jet optical luminosity is much lower than the luminosity of disc. We present the calculation of polarization degrees for a number of Seyfert-1 AGNs.
\end{abstract}


\section{Introduction}

The measurement of linear polarization of radiation from active galactic nuclei (AGN) presents additional information about the structure of  radiation sources and absorbing and scattering regions (the presence of jets, polar outflows, accretion discs near the AGN centres, the existence of dust toroidal clouds and  other clouds that scatter light). The polarimetric results are presented in many papers (see, for example,  Impey et al. 1991, Goodrich \& Miller 1994, Smith et al. 2002, 2004, 2005; Marin et al. 2012a,b; Marin \& Goosmann 2013; Marin 2014; Afanasiev et al. 2011, 2014 ). In these papers there are detailed references to the observational data and various models to explain these data. The important parameter in any explanation is the angle of inclination $i$ of the line of sight ${\bf n}$ relative to the normal ${\bf N}$ to equatorial plane of the system related to AGN. Note that the $i$ - value is determined by various methods which very frequently produce different results  (see the recent review of Marin 2016 ).

The inferences of the innermost structures of AGNs have been obtained indirectly by Rowan-Robinson (1977) suggested that AGNs are surrounded by dusty tori.  He also  gives the suggestion that Seyfert-2 galaxies are seen close to edge-on so that  the active nuclei are obscured by the tori.  Support for this picture came from  statistical investigation (see  Keel  1980) which demonstrated that Seyfert-1 galaxies are preferentially  seen face-on. The further importance of the orientation effects was considered in Lawrence \&  Elvis (1982) and De Zotti \& Gaskell (1985). Since then, the dusty - torus model has become the standard unified model.

When the optical polarization in continuum in AGNs was detected (see  Dibai \& Shakhovskoi 1966 and Walker 1966), it was initially accepted that this was the optical polarization of synchrotron radiation, since it had high intrinsic polarization. Later (Angel et al. 1976) the polarization of Balmer lines was discovered, which was similar to the continuum. After that it was clear that the scattering is responsible for both types of polarization.  A key point here was the lack of variability in polarization, which argues against a non-thermal synchrotron origin (see Becklin et al. 1973;  Knacke \& Capps (1974) for IR radiation and Antonucci \& Miller 1985; Bailey et al. 1988; Young et al. (1995) for spectropolarimetry).

According to  optical classification (see  Antonucci 1984, 1993;  Antonucci \& Miller 1985;  Lawrence 1987, 1993), there are
two types of Seyfert  AGNs  which also differ from each other by luminosities and spectra. While spectra of Seyfert-2 galaxies have a single set of relatively narrow emission lines, the spectra of Seyfert-1 galaxies have additional much broader components of hydrogen, helium and  other lines (see di Serego Alighieri et al. 1994, Lal et al. 2011). One of the main reasons for the difference between AGN-1 and AGN-2 is the different angle of observation.  Seyfert-1 corresponds to observation closer to the normal ${\bf N}$ to the equatorial plane.

It is the interesting feature that Seyfert-1 AGNs have mostly relatively small linear polarization $p\le 1\%$ and the inclination angle $ i\le 45^{\circ} -50^{\circ}$. On the contrary, for Seyfert-2 the polarization is larger ($p\ge 7\%$) and the inclination angle $ i\ge 60^{\circ}$ is higher (see Marin 2014). It is also important that the positional angle for Seyfert-1 corresponds to wave electric field oscillations  {\bf E} nearly parallel to the symmetry axis (more accurately parallel to the axis of the radio wave emission), and for Seyfert-2 the oscillations are parallel to the equatorial plane.

In the review of Marin (2014) various aspects are discussed: the different methods to estimate  the inclination angle $i$ and a number of  models to explain the observed polarimetric data.   Particularly, the 3-component model of AGN is discussed in detail. According to this model there are 3 components -  a conical polar outflow of scattering particles  with the half-opening angle $\sim 45^{\circ}$, a vast dust torus  in the equatorial plane and an optically thin ($\tau<1$)  equatorial flow  ( see also Smith et al. 2002; Goosmann \& Gaskell 2007).

According to this model, the scattering of non-polarized radiation from the AGN centre on electrons in a cone produces the  wave  electric field oscillations parallel to the equatorial plane. On the contrary, the scattering on electrons in the optically thin  equatorial flow produces the wave electric field  oscillations in the ''the line of sight - the axis of symmetry'' plane, that corresponds to  Seyfert-1 observations. The degree of this polarization is proportional to $\sin^2i$ and is not large. The second reason why Seyfert-1 AGNs have modest polarization degrees is the competition between these two scattering mechanisms with the mutually perpendicular position angles.

It is clear from a physical point of view that in axially symmetric models there can arise positional angles either perpendicular to the symmetry  axis or parallel to it. For this reason the competition can not explain the numerous cases of Seyfert AGNs with intermediate position angles. Such cases may be explained in models without axial symmetry (with the existence of separate scattering clouds, (see Goosmann  \& Matt 2011 and Marin et al. 2015)
, or in axially symmetric models by taking into account the Faraday rotation effect (see, for example, Silant'ev 2002;  Buliga et al. 2014) . Of course, both reasons may exist independently. The presence of magnetic field in jets and accretion discs is apparent in many objects (see Chakrabarti et al. 1994; Pariev \& Colgate 2007).

There are many types of jets in AGNs (see, for example,  Blandford 2003) both relativistic and non-relativistic ones.  Up to now various models of jets appear in literature (for example, Levinson \& Globus 2016). The  Faraday rotation of linearly polarized radiation in jets is studied in detail (Gomez et al. 2016). The core of jet sometimes is assumed to be optically thick  (Blandford \&  Konigle 1979; the review of Blandford 1990).

It is known that the base of jets ( hot corona) is the source of soft X-ray radiation  (see Haardt \& Maraschi 1993; Haardt \& Matt 1993). Far above this region the temperature diminishes and the optical radiation appears. Most frequently the adiabatic expansion of the medium is considered  (Blandford  2000, 2003). Recall that in adiabatic expansion the relation $T_1/T_2=(V_2/V_1)^{\gamma -1}$ takes place (Landau \& Lifshitz 1980). Here $T_1$ is the temperature in volume $V_1$, and $T_2$ is temperature in volume $V_2$. The parameter $\gamma=C_p/C_v$ is the ratio of thermal capacities at constant pressure and volume, respectively. For monoatomic gas $\gamma=5/3$ and for diatomic gas $\gamma=7/5$. Clear that expanding jet (mostly due to particles acceleration with the  distance from the centre which results in the increase of volume $V_2$ ) gives  rise to the inequality $V_2\gg V_1$ and consequently $T_2\ll T_1$. It is interesting that there are papers (for example, see Wang et al. 2005) where the existence of thermal matter in relativistic jets is discussed.

It should be noted that there are numerous Seyfert AGNs with intermediate values of polarization degrees and position angles (see, for example, Table 3 in Antonucci 1984; Table 4 in Smith et al. 2004). Note, that the aperture effects may cause distortions in both the measurement of polarization degree and the position angles of radiation in AGNs. In many  papers (see, for example, Lawrence 1991; Simpson 2005)  there exist the modifications of a standard model (the dependence of molecular and dust torus on luminosity of the central source, the particular distributions of dust grains in torus, and so on).

Usually one considers optically thick accretion discs (see Shakura \& Sunyaev 1973; Pariev \& Colgate 2007). It appears that both the optically thin and thick accretion discs really occur in AGNs.

\subsection{The statement of the problem}

Our objective is to discuss the mechanism for the origin of ''negative'' polarization by reflection of non-polarized radiation from a jet in optically thick accretion discs. The integral radiation from the surface of optically thick part of a jet  has small degree of polarization (see next Chapter). The thermal radiation from an optically thin jet is non-polarized. For this reason we consider only  reflection of non-polarized radiation in an accretion disc.

For the case of optically thin disc the problem is simple and corresponds to taking into account the single scattered radiation. When the incident radiation goes from the central nucleus of a disc the polarization degree of scattered radiation is described by formula (26). This formula corresponds to "negative" polarization.

The case of optically thick disc is more difficult for consideration. So, the optically thick accretion disc is illuminated by unpolarized radiation from a jet and we calculate the reflected radiation from the total surface of an accretion disc going into the  line of sight ${\bf n}$. We demonstrate that this radiation has "negative" linear polarization for all inclination angles $i$ between the line of sight ${\bf n}$ and the normal ${\bf N}$ to the accretion disc plane. Besides of reflected radiation, a telescope  detects the radiation directly going from the source. First we calculate this direct radiation from the optically thick jet (see Ch.2).

\begin{figure}
 \includegraphics[width=0.85\columnwidth]{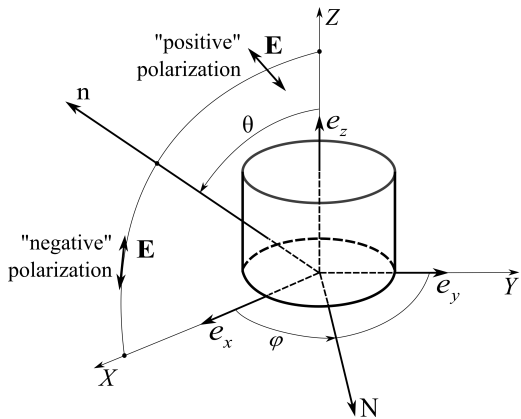}
 \caption{Polarization of radiation emerging from optically thick jet.}
\end{figure}

For calculating radiation fluxes  $F_I(\Theta)$ and Stokes parameter $F_Q(\Theta)$  (here  $\Theta$ is the angle between the line of sight ${\bf n}$ and direction of jet Z,  see Fig.1) we use the well known Chandrasekhar's (1950) results for the intensity $I(\mu)$ and the Stokes parameter $Q(\mu)$ in the Milne problem for atmospheres without absorption $\sigma_a/(\sigma_a+\sigma_s)\equiv q=0$. Here $\sigma_a$ and $\sigma_s$ are cross- sections of absorption and scattering, respectively. For calculating fluxes $F_I$ and $F_Q$ of radiation scattered in the optically thick accretion disc the results of the paper by Silant'ev \& Gnedin (2008) are used.

The calculations demonstrate that direct radiation from an optically thick jet for $ \Theta\le 60^{\circ}$ has the polarization with electric field oscillations perpendicular to the axis of a jet, and for $ \Theta\ge 60^{\circ}$ oscillations occur in the plane (${\bf n}Z$), where ${\bf n}$ is direction to an observer, and $Z$-axis directed along the jet. But this ''negative'' polarization is small $(p<0.2\%)$. Thus, polarization from the optically thick region of a jet for Seyfert-1 is positive, and for Seyfert-2 is ''negative'', just opposite to observations  (see, for example, Smith et al. 2002).

The radiation reflected  after multiple scatterings from an optically thick accretion disc  is ''negative'' for all angles of emission.  For inclination angles $i< 70^{\circ}$ this polarization is greater than that for the radiation escaping from an optically thick jet (the Milne problem, see Ch.3).

In conclusion we underline that we do not present a new model to explain the Seyfert -1 and 2 AGNs  features in full. Our objective is to call attention to a new mechanism of ''negative'' polarization formation in Seyfert-1 AGNs in optically thick accretion discs. Furthermore, we call attention to the explanation of intermediate position angles in these AGNs using the Faraday rotation effect. We hope that our calculations are applicable to all AGNs, not only to Seyfert AGNs.

\section{Polarization of radiation emerging from optically thick jet}

Let us introduce the necessary notations: the elemental radiating area  on the surface of a jet  is proportional to the element of the azimuthal angle $\varphi $: $dS=S_0d\varphi$. The axis Z is  directed along the jet, the direction to the telescope  ${\bf n}$ lies in the plane (XZ). The normal to the radiating  elemental area  is ${\bf N}$.  The  angle $\Theta$ is the angle between the Z-axis and the vector ${\bf n}$. The  vector  ${\bf n}$ is a linear combination of the coordinate unit vectors ${\bf e}_z$ and ${\bf e}_x$  (see Fig.1):

\begin{equation}
 {\bf n}={\bf e}_z\,\cos {\Theta} +{\bf e}_x\,\sin{\Theta}.
 \label{eq1}
\end{equation}
\noindent The normal ${\bf N}$ can be represented as

\begin{equation}
 {\bf N}={\bf e}_x\,\cos \varphi +{\bf e}_y\,\sin \varphi .
 \label{eq2}
\end{equation}

\noindent The cosine of angle between ${\bf n}$ and ${\bf N}$ is as follows:

\begin{equation}
 \mu={\bf nN}=\sin{ \Theta}\cos\varphi.
 \label{eq3}
\end{equation}

Chandrasekhar's (1950) intensity $I(\mu)$ [erg/cm$^2$s sr] and the Stokes parameter $Q(\mu)$ of the outgoing radiation can be approximated by the expressions:

\[
 I(\mu)=I(0)(1+2.18\mu-0.12\mu^2),
\]
\[
 Q(\mu)=I(0)(-11.71\%)\varphi_Q(\mu),
\]
\begin{equation}
 \varphi_Q(\mu)=(1-\mu^2)(1-1.52\mu+1.07\mu^2).
\label{eq4}
\end{equation}
\noindent Here $ I(0)= F/2\pi J_1 $ , where $ F$ is the total  flux of outgoing radiation; the value  $ J_1 $ is the first moment of the function in brackets. In our  approximation $ J_1 \simeq 1.257 $. The definition of various moments is given in Eq.(21) .  The ratio $Q(\mu)/I(\mu)$ determines the degree of linear polarization. The minus sign  in  $Q(\mu)$ implies that the wave electric vector ${\bf E}(\omega)$ oscillates perpendicular to the plane $({\bf nN})$. The Stokes parameter $U(\mu)=0$. Note that expressions (4) describe the Milne problem in conservative atmosphere with  $q=0$. Let us  also remind the reader that the Milne problem describes the situation when the sources of non-polarized radiation are located far from the surface of an atmosphere.

A telescope detects the radiation fluxes from areas with the azimuthal angles $\varphi$ in the range $-90^{\circ}<\varphi <90^{\circ}$:

\[
 F_I({\bf n})=\frac{S_0 }{R^2}\int_ {-\pi/2}^{ \pi/2}\, d\varphi \mu I(\mu),
\]
\[
 F_Q({\bf n})=\frac{S_0 }{R^2}\int_{ -\pi/2}^{\pi/2}\, d\varphi \mu
 Q(\mu)\cos2\varphi,
\]

\begin{equation}
 F_U({\bf n})=\frac{S_0 }{R^2}\int_{ -\pi/2}^{ \pi/2}\, d\varphi \mu
 Q(\mu)\sin2\varphi \equiv 0.
 \label{eq5}
\end{equation}

\noindent Here $R$ is the distance to the observed object. Expressions for $F_Q({\bf n})$ and $F_U({\bf n})$ take into account the transition of Stokes parameters into the telescope's coordinate system (XYZ). Using the expressions (4) we obtain:

\[
 F_I({\bf n})=\frac{2S_0 }{R^2}I(0)\sin{\Theta}(1+1.71\sin{\Theta}-0.08\sin^2{\Theta})
\]
\[
 \equiv \frac{1}{2 R^2}L_I(\Theta),
\]
\[
 F_Q({\bf n})=\frac{2S_0}{R^2}I(0)(-11.71\%)\sin \Theta\times
\]
\begin{equation}
 \left(\frac{1}{3}+0.028\sin^2{\Theta}-0.60 \sin{\Theta}\cos^2{\Theta}-0.41\sin^4{\Theta}\right).
 \label{eq6}
\end{equation}

\noindent The value 0.5$L_I(\Theta)$(erg/s) is the luminosity of observed part of a jet in the direction ${\bf n}$. Integrating $L_I(\Theta)$ over all directions of the angle $ \Theta$ (from $-\pi/2$ up to $\pi/2$) and taking into account that $L_I(\pi-\Theta)= L_I(\Theta)$, we obtain the total luminosity $L_0$ of radiating area along all directions:

\begin{equation}
 L_0=2\int_0^{\pi/2}d{\Theta} \,L_I(\Theta)=9.16  S_0I(0).
 \label{eq7}
\end{equation}

\noindent Using this expression, the formula for $L_I(\Theta)$ can be written in the form:

\begin{equation}
 L_I(\Theta)=\frac{L_0}{4.58}\sin {\Theta}(1+1.71\sin {\Theta}-0.08 \sin^2{\Theta}).
 \label{eq8}
\end{equation}

Note  that the positive sign in $F_Q({\bf n})$ corresponds to wave electric field oscillations in the plane  $({\bf ne_z})$ (see  Fig.1). This is the case of ''negative'' polarization in Seyfert-1 AGNs.

Table 1 and Fig.2 present the degrees of polarization  $p=F_Q({\bf  n})/F_I({\bf n})$  in percents for a number of angles $\Theta$. The second line of this Table presents the polarization in absorbing atmosphere with $q=0.1$. Here we approximated the values $I(\mu)$ and $Q(\mu)$ using the paper by Silant'ev (1980). Note that the calculation of angular distribution and polarization degree is connected with exact solution of characteristic equation and the  solution of the system of nonlinear equation for H-functions. This is difficult problem, especially for large values of absorption parameter $q > 0.5$. Some particular results are presented some later, before Eq.(24).

\begin{figure}
 \includegraphics[width=1.0\columnwidth]{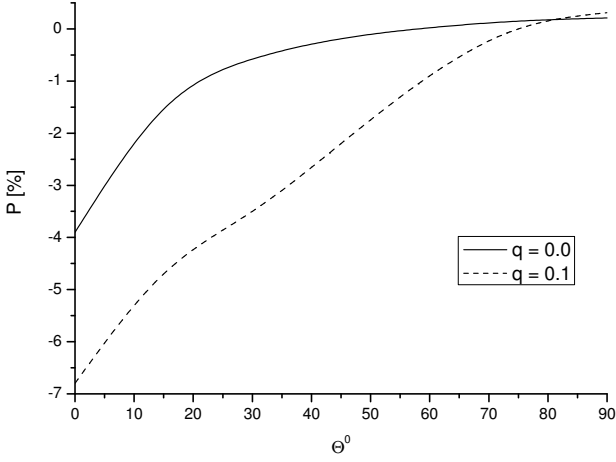}
 \caption{\small Polarization of radiation from an optically thick jet $p(\Theta)=F_Q(\Theta)/F_I(\Theta)$ [\%] for absorption degrees q = 0 and 0.1}
\end{figure}

\begin{table}
\centering
\caption{\small Polarization of radiation from an optically thick jet $p(\Theta)=F_Q(\Theta)/F_I(\Theta)$ [\%] for absorption degrees q = 0 and 0.1}
\begin{tabular}{p{0.5cm}p{0.5cm}p{0.7cm}p{0.7cm}p{0.7cm}p{0.5cm}p{0.5cm}p{0.5cm}}
\hline
$q \backslash \Theta^{\circ}$  & 0 & 15 & 30 & 45 & 60 & 75 & 90\\
\hline
$0$   &-3.9 & -1.54 & -0.58 & -0.19 & 0.02 & 0.15 & 0.21\\
$0.1$ &-6.8 & -4.7 & -3.5 & -2.2 & -0.9 & 0.00 & 0.31\\
\hline
\end{tabular}
\end{table}

\[
 I(\mu)=I_1(0)(1+2.63\mu-1.52\mu^2+2.29\mu^3),
\]
\begin{equation}
 Q(\mu)=I_1(0)(-20.4\%)(1-0.02\mu+1.05\mu^2-2.03\mu^3).
 \label{eq9}
\end{equation}
\noindent  The value $I_1(0)= F/2\pi J_1$ related with the ougoing flux $F$  by the same expression as in the case $q=0$. But in this case the first moment of function in brackets  is equal to $J_1\simeq 1.455$, i.e. the values $ I(0)$ and $I_1(0)$ are different.

In this case we obtain the expressions:
\[
 F_I({\bf n})=\frac{2S_0 }{R^2}I_1(0)\sin {\Theta}(1+2.065\sin{\Theta}-1.013\sin^2{\Theta}+
\]
\[
 1.348\sin^3{\Theta})\equiv \frac{1}{2R^2}\,L_I({\Theta}),
\]
\[
 F_Q({\bf n})=\frac{2S_0 }{R^2}I_1(0)(-20.4\%)\sin{\Theta}
 \left(\frac{1}{3}-0.008\sin {\Theta}+\right.
\]
\begin{equation}
 \left.0.42\sin^2{\Theta}-0.8 \sin^3{\Theta}\right).
 \label{eq10}
\end{equation}

\noindent According to our calculations, the ''negative'' polarization occurs at  $\Theta>(60-75^{\circ})$. This does not correspond to observation of Seyfert-1 ( e.g. Smith et al. 2002).

If  the  magnetic field ${\bf B}$ exists in a jet, then due to Faraday rotation of the polarization plane (the rotation angle $\Psi=0.4\lambda^2${\bf nB}$\tau\equiv \delta({\bf nB})\tau/2$, where $\tau$ is the optical Thomson length, $\lambda$ is the wavelength in microns, and the magnetic field in Gauss)  the Stokes parameter $F_U $ arises. Note that the positional angle of polarization $\chi$ is related to  parameters  $F_Q$ and $F_U$ : $\tan2\chi=  F_U/F_Q$. The absorption factor in the radiative transfer equation for the value $Q(\tau,\mu)+iU(\tau,\mu)$ is of the form $(1+i\delta ({\bf nB}))$ . Because the radiation polarization  arises mainly  at the last scattering  before  escaping from atmosphere, a good approximation for the values  $F_Q({\bf n},{\bf B})$ and $ F_U({\bf n},{\bf B})$  are the following formulas  ( see Silant'ev  2002, 2007; Silant'ev et al. 2009, 2013):

\[
 F_Q({\bf n},{\bf B})=\frac{S_0 }{R^2}I(0)(-11.71\%)\sin{\Theta}\times
\]
\[
 \int_{-\pi/2}^{\pi/2}\,d\varphi\,\cos\varphi\cos2\varphi
 \frac{\varphi_Q(\mu)} {1+\delta^2(\Theta,\varphi)},
\]
\[
 F_U({\bf n},{\bf B})=\frac{S_0 }{R^2}I(0)(-11.71\%) \sin{\Theta}\times
\]

\begin{equation}
 \int_{-\pi/2}^{\pi/2}\,d\varphi\,\cos \varphi\sin2\varphi
 \frac{\varphi_Q(\mu)\delta(\Theta,\varphi)}{1+\delta^2(\Theta,\varphi)}.
 \label{eq11}
\end{equation}

\noindent  In deriving equations (11) we used the relation (3) in the explicit form. The function $\delta(\Theta,\varphi)$ has the form (we use the cylindrical reference frame):

\[
 \delta(\Theta,\varphi)\equiv 0.8\lambda^2{\bf nB}=
\]
\begin{equation}
 0.8\lambda^2\,(B_z\cos \Theta+B_{\rho}\sin \Theta\cos\varphi -B_{\varphi}\sin \Theta\sin\varphi).
 \label{eq12}
\end{equation}

In the case of the vertical magnetic field $B_z$ the value  $\delta(\Theta,\varphi)$ does not depend on $\varphi$ and the polarization  degree  of emerging radiation becomes:

\begin{equation}
 p({\bf nB})=\frac{\sqrt{F_Q^2+F_U^2}}{F_I}=\frac{p_{rel}}{\sqrt{1+(0.8\lambda^2B_z\cos{\Theta})^2}}.
 \label{eq13}
\end{equation}
\noindent Here the degree of polarization  $p_{rel}$  is the value when in the absence of the magnetic field (see  Chandrasekhar 1950). Note that practically full depolarization of optical radiation with $\lambda \simeq 0.5 $ microns occurs at $B_z\simeq 10 $ Gauss.

In the case of the vertical magnetic field ${\bf B}=B_z{\bf e_z}$ the relation  $ F_U({\bf n},{\bf B}) \equiv 0$ occurs, i.e. the wave electric field oscillations may be parallel to axis of jet or perpendicular to it.

It is known ( see, for example, Landau \& Lifshitz 1984) that the Faraday rotation of separate beam of light does not change the degree of polarization. At multiple light scattering we summarize the fluxes with different Faraday rotations. This leads to depolarization of emerging radiation, so called the Faraday  depolarization. Eq.(13) describes this effect.

In the presence of  the azimuthal magnetic field  $\delta(\Theta,-\varphi)\neq \delta(\Theta,\varphi)$, and the Stokes parameter $F_U({\bf n},{\bf B})\neq 0$. It means that the position angle $\chi $ ($\tan2\chi=F_U/F_Q$) does not take the values $0^{\circ}$ or $90^{\circ}$. The observed positional angle lies between these boundary values. The largest value of the angle $\chi$ is equal to $45^{\circ}$. This value corresponds to $\delta=0.8\lambda^2B_{\varphi}\gg 1$. Thus, the existence of the azimuthal magnetic field  $ B_{\varphi}$ in a jet may explain the intermediate values of the positional angle.

\section{Polarization of radiation reflected from an optically thick accretion disc} 

Non-polarized radiation reflected from an optically thin accretion disc has  large polarization degree (see, formula (26) in this chapter). Positional angle of the reflected radiation is parallel to the normal  ${\bf N}$ to the disc, i.e. this is ''negative'' polarization. Usually the accretion disc is considered as optically thick. The radiation emerging from such disc (the Milne problem) has the wave electric oscillations {\bf E} parallel to disc's plane.  We demonstrate below  that reflection of non-polarized light from a jet in the optically thick disc leads to ''negative'' polarization for all  angles between the line of sight ${\bf n}$ and the normal ${\bf N}$ to the disc. The degree of reflected radiation is smaller than that  in the case of an optically thin disc but is significantly large: $p_{max}\simeq 8.4\%$ for the case of an optically thick jet, and $p_{max}\simeq 4.1\%$ for an optically thin jet.  These estimates refer to conservative atmosphere with $q=0$.

The optical radiation from  a jet can be considered as an effective point-like source of non-polarized radiation. It is not important whether the radiating region in a jet is optically thin or thick.  It is known (see  Grinin \& Domke 1971; Silant'ev \& Gnedin 2008) that reflected radiation does not depend on the height of a point source above the plane of the optically thick medium.   Let us explain this effect. It is  known that the illumination of plane-parallel atmosphere by a point-like source gives rise to "spot-like" distribution of radiation inside the atmosphere. In this case  the free term of the radiative transfer equation has the form :
\[
S(z,\rho,\varphi)\sim \frac{\exp{(-\tau/\mu')}}{\rho^2+(z+h)^2},
\]
\noindent where  $\tau$ is vertical ( along Z-axis) optical depth, $h$ is the height of a point-like source over the plane of an atmosphere, $ \vartheta' (\mu'=\cos\vartheta')$ characterizes the direction of  incident radiation ($\mu'=(z+h)/\sqrt{\rho^2+(z+h)^2})$. In this case the integration of the radiative transfer equation over the radial variable $\rho d\rho$  leads to the usual radiative transfer equation depending on the vertical optical depth $\tau$  and on the direction ${\bf n}$ to the telescope. Note that the integration over $\rho d\rho$ corresponds to summation the fluxes $F_I, F_Q$ from the all "spot-like" distribution of radiation in the disc.  The integration over $\rho d\rho$ in the interval $(0,\infty)$ can be substituted by integration over $d\mu'/\mu'$ in the interval (0,1). Here we  use the relation $\rho d\rho=(\rho^2+(z+h)^2)d\mu'/\mu'$. The term $(\rho^2+(z+h)^2)$  cancels the same term in the source $S(z,\rho,\varphi)$ and the problem becomes independent of height $h$. In Fig.3 we present schematically the geometry of the origin of polarized radiation -  scattered from a point-like source and polarized radiation emerging from deep layers of optically thick accretion disc (the Milne problem). Note that the independence on height occurs if the scattered atmosphere is plane-parallel and the source is point-like. This means  that the sum of point-like sources located in a line also can be described by our formulas.

\begin{figure}
 \includegraphics[width=0.75\columnwidth]{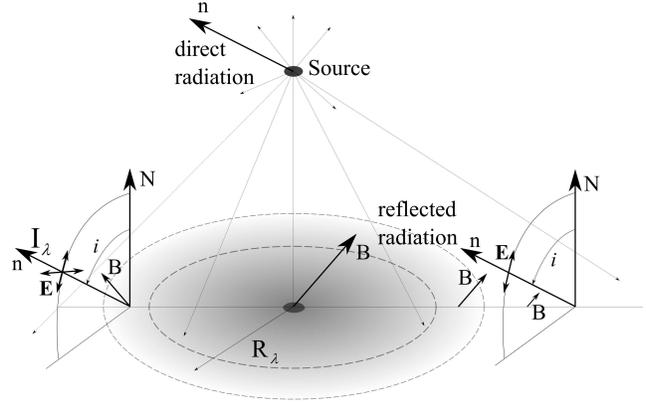}
 \caption{Reflected radiation from source and the Milne problem radiation $I_\lambda$.}
\end{figure}

According to Silant'ev \& Gnedin (2008) the radiation flux scattered in an optically thick atmosphere can be presented in the form  ($\mu={\bf nN}\equiv \cos i)$:

\[
 F_I(\mu)=\frac{3}{2 R^2}(1-q)\mu\,\int_0^1\frac{dx}{\mu+x}L_I(x)[2A(\mu)A(x)+
\]
\[
 B(\mu)B(x)],
\]
\[
 F_Q(\mu)=\frac{3}{2 R^2}(1-q)\mu\,\int_0^1\frac{dx}{\mu+x}L_I(x)[2C(\mu)A(x)-
\]
\begin{equation}
 D(\mu)B(x)].
 \label{eq14}
\end{equation}

\noindent Recall that  $i$ is the angle between the direction ${\bf n}$ to an observer and the normal to the accretion disc ${\bf N}$, $R$ is the distance to the telescope, $q$ is the degree of light absorption in an atmosphere. If the direction of the jet is perpendicular to the disc's plane, then the angle $\Theta$ in the previous chapter coincides with angle $i$. We consider this case. So, the expression for luminosity $L_I(\mu)$ is presented by Eq.(8).

The functions $A(\mu), B(\mu), C(\mu)$ and $D(\mu)$ satisfy the system of nonlinear equations similar to Chandrasekhar's (1950) $H(\mu)$-functions:

\[
 A(\mu)=\frac{1+\mu^2}{4}+\frac{3}{8}(1-q)\mu\,\int_0^1\frac{dx}{\mu+x}\times
\]
\[
 [(1+x^2)(2A(\mu)A(x)+B(\mu)B(x))+
\]
\[
 (1-x^2)(2A(\mu)C(x)-B(\mu)D(x))],
\]

\[
 B(\mu)=\frac{1-\mu^2}{2}+\frac{3}{4}(1-q)\mu\,\int_0^1\frac{dx}{\mu+x}\times
\]
\[
 (1-x^2)[(2A(\mu)A(x)+B(\mu)B(x))-
\]
\[
 (2A(\mu)C(x)-B(\mu)D(x))],
\]
\[
 C(\mu)=\frac{1-\mu^2}{4}+\frac{3}{8}(1-q)\mu\,\int_0^1\frac{dx}{\mu+x}\times
\]
\[
 [(1+x^2)(2C(\mu)A(x)-D(\mu)B(x))-
\]
\[
 (1-x^2)(2C(\mu)C(x)+D(\mu)D(x))],
\]
\[
 D(\mu)=\frac{1-\mu^2}{2}+\frac{3}{4}(1-q)\mu\,\int_0^1\frac{dx}{\mu+x}\times
\]
\[
 (1-x^2)[(2C(\mu)A(x)-D(\mu)B(x))-
\]
\begin{equation}
 (2C(\mu)C(x)+D(\mu)D(x))].
 \label{eq15}
\end{equation}

\noindent The sum of  the first two equations of this system gives the relation:

\[
 \frac{3}{2}(1-q)\mu\,\int_0^1\frac{dx}{\mu+x}[2A(\mu)A(x)+B(\mu)B(x)]=
\]
\begin{equation}
 2A(\mu)+B(\mu)-1.
 \label{eq16}
\end{equation}

\noindent Subtraction of the third equation from the fourth one gives the other relation:

\[
 \frac{3}{2}(1-q)\mu\,\int_0^1\frac{dx}{\mu+x}[2C(\mu)A(x)-D(\mu)B(x)]=
\]
\begin{equation}
 2C(\mu)-D(\mu).
 \label{eq17}
\end{equation}

Let us approximate  expression (8) for the luminosity $L_I(\mu)$ as a power series:

\[
 L_I(\mu)=S_0I(0)\,2.63\Phi_1(\mu),
\]
\begin{equation}
 \Phi_1(\mu)=a_0+a_1\mu+a_2\mu^2+a_3\mu^3+a_4\mu^4,
 \label{eq18}
\end{equation}

\noindent where the coefficients $a_n$ have the explicit form:

\begin{equation}
 a_0=1, \,\,a_1=-0.06,\,\,a_2=-0.58,\,\,a_3=0.06,\,\,a_4=-0.42.
 \label{eq19}
\end{equation}

Substitution of Eq. (18) in Eq. (14) and taking into account the relations (16) and (17) leads to the following expressions for the radiation  fluxes $F_I(\mu)$ and $F_Q(\mu)$ reflected from the accretion disc :

\[
 F_I(\mu)=\frac{S_0\,I(0)}{R^2}\,2.63\,[(2A(\mu)+B(\mu)-1)\Phi_0(\mu)+
\]
\[
 (2A(\mu)A_0+B(\mu)B_0)f_0(\mu)+
\]
\[
 +(2A(\mu)A_1+B(\mu)B_1)f_1(\mu)+(2A(\mu)A_2+
\]
\[
 B(\mu)B_2)f_2(\mu)+
 (2A(\mu)A_3+B(\mu)B_3)f_3(\mu)],
\]
\[
 F_Q(\mu)=-\frac{S_0\,I(0)}{R^2}\,2.63\,[(2C(\mu)-D(\mu))\Phi_0(\mu)+
\]
\[
 (2C(\mu)A_0-D(\mu)B_0)f_0(\mu)+
\]
\[
 +(2C(\mu)A_1-D(\mu)B_1)f_1(\mu)+(2C(\mu)A_2-D(\mu)B_2)f_2(\mu)+
\]
\begin{equation}
 (2C(\mu)A_3-D(\mu)B_3)f_3(\mu)].
 \label{eq20}
\end{equation}

\noindent Here $A_n,\,B_n,\,C_n$ and $D_n$ are the moments of functions $A(\mu), B(\mu), C(\mu)$ and $D(\mu)$:

\begin{equation}
 A_n=\int_0^1\,d\mu\,\mu^n\,A(\mu).
 \label{eq21}
\end{equation}

\noindent Analogous expressions can be derived for the values $B_n,\,C_n,\,D_n$.

Functions $\Phi_0(\mu),f_n(\mu)$ have the following form:
\[
 \Phi_0(\mu)=a_0-a_1\mu+a_2\mu^2-a_3\mu^3+a_4\mu^4,
\]
\[
 f_0(\mu)=\frac{3}{2}(1-q)\mu(a_1-a_2\mu+a_3\mu^2-a_4\mu^3),
\]
\begin{equation}
 f_1(\mu)=\frac{3}{2}(1-q)\mu(a_2-a_3\mu+a_4\mu^2),
 \label{eq22}
\end{equation}
\[
 f_2(\mu)=\frac{3}{2}(1-q)\mu(a_3-a_4\mu),
\]
\[
 f_3(\mu)=\frac{3}{2}(1-q)\mu\,a_4.
\]

Hereafter we restrict ourselves to the cases of absence of light absorption  ($q=0$)  and the nonconservative  atmosphere with $q=0.1$. The functions  $A(\mu), B(\mu), C(\mu)$ and $D(\mu)$ for these cases are presented in Tables 2 and 3.  The moments  $A_n,\,B_n,\,C_n$ and $D_n$ are given in Table 4. Note that at  $q=0$ the relation (16) allows us to  obtain the quadratic  relation between the moments $A_0$ and $B_0$. Considering this relation as quadratic equation for $A_0$ (or for $B_0$) we can obtain  the exact values for moments $A_0=2/3$ and $B_0=2/3$. In this case the term with $f_0(\mu)$ acquires simpler form.

It should be noted that in expressions for $F_I(\mu)$  and $F_Q(\mu)$  in Eq.(20) we are adding the fluxes of radiation, which directly, without scattering, goes to the telescope. For an optically thick jet  this flux is due to half of the luminosity. This  means that within the parenthesis of expression for $F_I(\mu)$  in Eqs.(20) we can substitute the term $0.5\Phi_1(\mu)$. In brackets for $F_Q(\mu)$ we are to add the term ($-0.5\Phi_1(\mu)p(\mu)$), where the polarization degree $p(\mu)$  is determined from Table 1. The approximation for $p(\mu)$ has the form:

\begin{equation}
 p(\mu)\simeq 0.01(0.21+2.68\mu-20.51\mu^2+43.52\mu^3-29.8\mu^4).
 \label{eq23}
\end{equation}

\begin{table}
\caption{The functions  $A(\mu), B(\mu), C(\mu), D(\mu)$, and polarization degrees in percents  $p_{ref}(\mu), p_{point}(\mu), p_{rel}(\mu)$, and the angular distribution $J_{rel}(\mu)$ of radiation  at q=0}
\begin{tabular}{|p{0.3cm}|p{0.6cm}|p{0.6cm}|p{0.6cm}|p{0.6cm}|p{0.4cm}|p{0.4cm}|p{0.8cm}|p{0.4cm}|}
\hline
$\mu $  & A      & B      & C      & D      & $p_{ref}$ & $p_{point}$ & $p_{rel}$ &$ J_{rel}$ \\
\hline
  0     & 0.25   & 0.5    & 0.25   & 0.5    & 0.21 & 0    & -11.71 & 1 \\
  0.05  & 0.2860 & 0.5739 & 0.2657 & 0.5557 & 4.64 & 2.12 & -8.99  & 1.146 \\
  0.1   & 0.3195 & 0.6237 & 0.2737 & 0.5860 & 6.43 & 3.09 & -7.45  & 1.264 \\
  0.2   & 0.3890 & 0.6948 & 0.2794 & 0.6163 & 8.00 & 3.87 & -5.41  & 1.483 \\
  0.3   & 0.4645 & 0.7385 & 0.2745 & 0.6173 & 8.36 & 4.08 & -4.04  & 1.690 \\
  0.4   & 0.5469 & 0.7587 & 0.2602 & 0.5936 & 7.98 & 3.94 & -3.03  & 1.892 \\
  0.5   & 0.6365 & 0.7571 & 0.2373 & 0.5473 & 7.40 & 3.60 & -2.25  & 2.091 \\
  0.6   & 0.7335 & 0.7345 & 0.2060 & 0.4793 & 6.55 & 3.04 & -1.63  & 2.287 \\
  0.7   & 0.8380 & 0.6913 & 0.1666 & 0.3902 & 5.30 & 2.41 & -1.11  & 2.483 \\
  0.8   & 0.9500 & 0.6279 & 0.1190 & 0.2805 & 3.65 & 1.68 & -0.68  & 2.677 \\
  0.9   & 1.0696 & 0.5443 & 0.0635 & 0.1503 & 1.90 & 0.87 & -0.32  & 2.870 \\
  0.95  & 1.1322 & 0.4951 & 0.0327 & 0.0777 & 0.98 & 0.45 & -0.15  & 2.967 \\
  1     & 1.1968 & 0.4408 & 0      & 0      & 0    & 0    & 0      & 3.063 \\
\hline
\end{tabular}
\end{table}

In Tables 2 and 3  we present the following values of degrees of polarization: $p_{ref}(\mu)=F_Q(\mu)/F_I(\mu)$ is the polarization of jet radiation, reflected from the optically thick accretion disc;  $p_{point}(\mu)$  is the radiation polarization of a point - like source of isotropic non-polarized radiation reflected from the disc;  $p_{rel}(\mu)$ is the polarization of radiation, emerging from the accretion disc in the case of Milne's problem (the sources are located far from the surface of atmosphere). Note once more that the minus sign denotes the wave electric field oscillations perpendicular to the plane  $({\bf nN})$ (the so called positive polarization). In the last column in Table 2 we present the angular distribution $J_{rel}(\mu)=I(\mu)/I(0)$ of the radiation in the Milne problem (see  Chandrasekhar 1950).

Note that  in nonconservative atmosphere ($q\neq 0$) the value of angular distribution $J_{rel}(\mu)$ became higher along the normal ${\bf  N}$ with increasing of absorption value $q$. So, for $\mu=0.9$ and $1$, and $q=0.2,0.3, 0.5$ the values  $J_{rel}(\mu)$ are $5.02, 6.49; 6.65, 10.05;11.35, 31.7$, respectively. The polarization degree became  larger near direction perpendicular to ${\bf N}$. For $\mu=0$ and $0.1$, and the same $q$-values we have $p=28.6\%, 25\%; 36.6\%, 33.4\%; 52.7\%, 50.0\%$ (see  Silant'ev 1980, where the results up to $q=0.5$ are presented). Recall  that the electric field ${\bf E}$ oscillations are parallel to the plane of atmosphere. The increasing of angular distribution at $\mu\to 1$ and the polarization degree at $\mu\to 0$ is the consequence that radiation absorbs greater if it has larger path before escaping from the surface. As a result, for high absorption the picture looks like single scattering of radiation before escaping the atmosphere.

Anisotropic luminosity $L_I(\mu)$ (see  Eq.(8), where $\Theta= i$ for considered case  with the jet perpendicular to the disc) increases the degree of polarization $p_{ref}(\mu)$ to approximately twice $p_{point}(\mu)$ for isotropic  luminosity with $L_I(\mu)=L_0/4\pi$, which corresponds to the case of an optically thin  jet:

\[
 F_I(\mu)=\frac{L_0}{4\pi R^2}[2A(\mu)+B(\mu)]
\]
\begin{equation}
 F_Q(\mu)=-\frac{L_0}{4\pi R^2}[2C(\mu)-D(\mu)]
 \label{eq24}
\end{equation}

The expression for $F_I(\mu)$ in this formula also takes into account the radiation directly propagated to the telescope from the optically thin jet. The expression  $ 2C(\mu)-D(\mu)$ is negative for all values of $\mu$. This means that the polarization of reflected radiation is always ''negative''.  Note also that according to Eqs.(15) functions $A, B, C $ and $D$ depend on the degree of absorption $q$. The angular distribution $J_{point}(\mu)= F_I(\mu)/F_I(0)\equiv 2A(\mu,q)+B(\mu,q)$.
It is of interest that the polarization degree $p_{ref}(\mu,q=0)$ in the angular range  $i< 80^{\circ}(\mu>0.2)$ is greater in absolute value than the polarization degree $p_{rel}(\mu, q=0)$. The inequality $p_{point}(\mu,q=0)>|p_{rel}(\mu,q=0)|$ occurs for $i< 70^{\circ}(\mu>0.3$). The total flux of radiation detected by an observer is equal to $F_{tot}(\mu)=F_{jet}(\mu)+F_{rel}(\mu)$, where $F_{jet}(\mu)$ consists of both direct radiation from a jet and after multiple scatterings in the disc. The flux $F_{rel}(\mu)$ emerges from the ring surface of the disc with a central radius (see  Poindexter et al. 2008):

\begin{table}
\caption{The functions  $A(\mu), B(\mu), C(\mu), D(\mu)$, and polarization degrees in percents   $p_{point}(\mu), p_{rel}(\mu)$, and the angular distributions $J_{point}(\mu)$ and $J_{rel}(\mu)$ of radiation  at q=0.1}
\begin{tabular}{|p{0.3cm}|p{0.6cm}|p{0.6cm}|p{0.6cm}|p{0.6cm}|p{0.4cm}|p{0.4cm}|p{0.8cm}|p{0.4cm}|}
\hline
$\mu $  & A      & B      & C      & D      & $p_{point}$ & $J_{point}$ & $p_{rel}$ & $J_{rel}$ \\
\hline
  0     & 0.25   & 0.5    & 0.25   & 0.5    & 0 & 1    & -20.36 & 1 \\
  0.05  & 0.2749 & 0.5577 & 0.2635 & 1.89 & 1.11 & 1.107 & -17.88  & 1.135 \\
  0.1   & 0.2961 & 0.5921 & 0.2702 & 0.5726 & 2.72 & 1.184 & -16.39 & 1.249 \\
  0.2   & 0.3381 & 0.6323 & 0.2742 & 0.5940 & 3.49 & 1.308 & -14.18  & 1.470 \\
  0.3   & 0.3823 & 0.6447 & 0.2684 & 0.5888 & 3.69& 1.409& -12.35  & 1.698 \\
  0.4   & 0.4301 & 0.6338 & 0.2540& 0.5615 & 3.58& 1.494& -10.66  & 1.944\\
  0.5   & 0.4825 & 0.6017 & 0.2315 & 0.5141 & 3.27 & 1.567 & -9.00  & 2.216 \\
  0.6   & 0.5399 & 0.5493 & 0.2009 & 0.4476& 2.81& 1.629 & -7.31  & 2.524 \\
  0.7   & 0.6028 & 0.4775 & 0.1624 & 0.3625 & 2.24 & 1.683 & -5.58 & 2.878 \\
  0.8   & 0.6715 & 0.3867 & 0.1161 & 0.2594 & 1.58 & 1.730 & -3.78  & 3.293 \\
  0.9   & 0.7462 & 0.2773 & 0.0619 & 0.1385 & 0.83 & 1.770 & -1.92  & 3.789 \\
  0.95  & 0.7859 & 0.2157 & 0.0319 & 0.0714 & 0.42 & 1.787 & -0.97  & 4.075 \\
  1     & 0.8271 & 0.1495 & 0      & 0      & 0    & 1.804   & 0      & 4.392 \\
\hline
\end{tabular}
\end{table}

\[
 R_{\lambda}\simeq 0.97\times 10^{10}\left(\frac{\lambda_{rest}}{1\mu m}\right)^{4/3}\times
\]
\begin{equation}
 \times \left(\frac{M_{BH}}{M_{\odot}}\right)^{2/3}
 \left(\frac{L_{bol}}{\varepsilon L_{Edd}}\right)^{1/3} cm.
 \label{eq25}
\end{equation}

\noindent Here $L_{Edd} $ is the Eddington luminosity, $L_{bol}=\varepsilon \dot M c^2$ is the bolometric luminosity, $\varepsilon$ is the conversion factor of the gravitational energy of the accreting gas to radiation, $c$ is speed of light and $\dot M$ is the accretion rate. This ring surface we denoted by dotted lines in Fig.3.

Competition between two mechanisms  $p_{ref}(\mu)$ (or $p_{point}(\mu)$) and  $p_{rel}(\mu)$ with mutually perpendicular position angles leads at $q=0$ to the observed ''negative'' polarization $0.5\%$ at $\mu=0.6$  ($i=53^{\circ}$), if the luminosity $L_{jet}(\mu =0.6)=0.26 L_{tot}(\mu=0.6)$. For the optically thin jet the same ratio occurs at larger jet luminosity  $L_{jet}(\mu =0.6)=0.46 L_{tot}(\mu=0.6)$. For polarization of $0.2\%$ at $\mu=0.9$ ($i=26^{\circ}$) the ratios are  practically the same. We see that observing ''negative'' polarization may be possible even at smaller jet luminosities as compared to the luminosity of the disc.  For an absorbing atmosphere it is more difficult to obtain
''negative'' polarization in observed radiation (see Table 3).

The presence of magnetic field with ($\delta ({\bf nB})\ge 10$, that gives for $\lambda \simeq 0.5$ microns the value of magnetic field $B\simeq 10$ Gauss) within the ring with radius  $R_{\lambda}$ diminishes the polarization due to Faraday depolarization of Milne's radiation emerging from the accretion disc and  the part of radiation falling into the circle with radius  $R_{\lambda}$ from the jet (see Fig.3). But the reflected jet radiation outside this ring  does not undergo Faraday depolarization (we assume that magnetic field diminishes with increasing distance from the centre). In this situation the only polarized radiation is the reflected jet emission. For this reason, the observed ''negative'' polarization in Seyfert-1 AGN occurs at smaller jet luminosity. So, the ''negative'' polarization of $0.5\%$ at $\mu=0.6$ requires the ratio $L_{jet}/L_{tot} \simeq 0.08$ for an optically thick jet, and the value of $0.16$ in the case of the optically thin jet. The corresponding values for polarization of $0.2\%$ at $\mu=0.9$ are equal to $0.11$ and $0.23$. The values for $p_{ref}(\mu)$ and $p_{point}(\mu)$ have to be slightly less then in Table 2 because the reflected radiation beyond the ring $R_{\lambda}$ is more polarized than in the region with $R<R_{\lambda}$ ( this part of radiation  is closer to the case of singly scattered radiation, see  below).

Let us consider qualitatively why the ''negative'' polarization arises during scattering in the optically thick atmosphere, i.e. the wave electric vector {\bf E} oscillates in the plane $({\bf nN})$. First let us recall that scattering of non-polarized radiation  gives rise to oscillations perpendicular to the plane with normal $({\bf nN})$. If the angle of scattering is  $90^{\circ}$, then the scattered radiation is $100\%$ polarized (see Fig.4).

\begin{figure}
 \includegraphics[width=0.85\columnwidth]{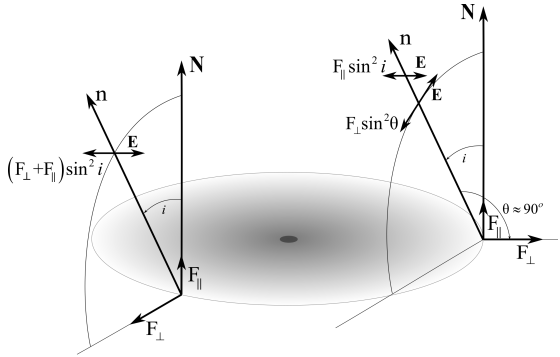}
 \caption{Origin of ''negative'' polarization.}
\end{figure}

An incident light beam produces in the atmosphere the distribution of radiation density with the maximum at the center of the beam (''spot''). The radiation diffuses both along the  normal  ${\bf N}$ and radially,  perpendicular to  ${\bf N}$. The first flux,  $F_{\parallel}$, gives rise to the usual polarization of the emerging light $\sim \sin^2i$  with  oscillations perpendicular to $({\bf nN })$.

The second flux,  $F_{\perp}$, is polarized perpendicular to the plane  $({\bf nN})$, if it is directed to an observer or in the opposite direction. If the flux $F_{\perp}$ is directed perpendicular to the plane $({\bf nN})$, then the polarization in this plane appears. But in this case the scattering angle does not coincide with the angle  $i$ and is near  $90^{\circ}$, i. e.  the polarization degree is near  $100\%$. Thus, the ''negative'' polarization arises if  $F_{\parallel}< F_{\perp}$. In the Milne problem $F_{\perp}=0$ and emerging radiation has wave electric oscillations parallel to the surface of the accretion disc.

The reason why the  anisotropy of the luminosity of source $L_I(\mu)$ leads to larger polarization  of reflected radiation as compared to the case of isotropic luminosity is clear. The anisotropy enhances the part of the radiation falling far from the center of the ''spot'', where the flux $F_{\perp}$ is more pronounced in comparison with $F_{\parallel}$.

As an example when  $F_{\parallel}=0$, we consider the single scattering of non-polarized radiation from a point-like source in a optically thin accretion disc. The source is placed at the center of the disc.  In this case the radiation inside the disc falls off as  $\exp{(-\tau_{\perp})}$, where  $\tau_{\perp}$ is radial optical depth in the disc. Integrating over the whole surface of the disc, we obtain the degree of polarization from the radiating ''spot'':

\begin{equation}
 p(\mu)=\frac{F_Q(\mu)}{F_I(\mu)}=\frac{\sin^2\vartheta}{1+\cos^2\vartheta+2\sin^2\vartheta},
 \label{eq26}
\end{equation}

\noindent where $\vartheta\equiv i$ is the angle between the normal to the disc ${\bf N}$ and the line of sight ${\bf n}$.

The positive sign of this expression implies that wave electric field oscillations occur in the plane  $({\bf nN})$, i.e. this is ''negative'' polarization. Maximum value of polarization of $33\%$ corresponds to the angle $\vartheta=90^{\circ}$. Polarization (26) reaches maximum during the scattering of non-polarized radiation in the optically thin accretion disc. Clear multiple scatterings only decrease polarization. If the radiation escapes along normal  ${\bf N}$ ($\mu=\cos i=1$) the polarization is absent due to axial symmetry. It should be noted that polarization (26) takes place  for radiation escaping from  every circular ring of the optically thin accretion disc.

The mechanism for the origin of ''negative'' polarization is considered in the book of Dolginov et al. (1995) in more detail.

\begin{table}
\centering
\caption{\small The moments $A_n$, $B_n$, $C_n$, $D_n$ for  $q=0$ (the upper Table) and $q=0.1$ (the lower Table)}
\begin{tabular}{|r|cccc}
\hline
 n & $A_n$ & $B_n$ & $C_n$ & $D_n$ \\
\hline
       0   &0.66667 &0.66667 &0.20146 &0.45372\\
       1  &0.41151 &0.32522 &0.07886 &0.18161\\
       2  &0.30238 &0.20795 &0.04285 &0.09962\\
       3  &0.24001 &0.15022 &0.02707 &0.06324\\
       4  &0.19930 &0.11645 &0.01868 &0.04380\\
\hline
\hline
       0 &0.50261 &0.51448 &0.19732 &0.43128 \\
       1 &0.29817 &0.22454 &0.07701 &0.17054 \\
       2 &0.21571 &0.13285 &0.04182 &0.09301 \\
       3 &0.16983 &0.09038 &0.02640 &0.05885\\
       4 &0.14033 &0.06674 &0.01823 &0.04066\\
\hline
\end{tabular}
\end{table}

\subsection{Calculation of ''negative'' polarization for a number of Seyfert-1 AGNs}

To demonstrate our mechanism of ''negative'' polarization in  Seyfert-1 AGNs  we  considered some objects  with measured   polarization degrees $p(i)$ and optical  luminosity  $L_{disc}$ (see  Ref (1) - Smith et al. 2002; Ref (2) - Martin et al. 1983 ; Ref (3)-  Berriman 1989).  A crucial part of this calculation is computing the inclination angles for these objects.  The values of inclination angles can be obtained from the virial parameters that define the geometry, velocity field and orientation of broad line regions (BLR). The expression for the virial parameter is used to determine the mass of a supermassive black hole (SMBH) at the centre of an AGN.

There are various approaches for determining the value of the virial parameter. A disc like geometry for the BLR has been proposed by several authors (Labita et al. 2006; Decarli et al. 2008; Kollatschny et al. 2006; Kollatschny \& Zetzl 2013). McLure \& Jarvis (2002) have shown that for a disc inclined at the angle $i$ to the observer the virial coefficient value is:

\begin{equation}
f=\frac{1}{4\sin^2 i}.
\label{eq27}
\end{equation}
\noindent Using this expression it is possible to obtain  from the virial theorem the following formula for $\sin i$ ( Piotrovich et al. 2015):

\begin{equation}
\sin i=\frac{1}{2}\left(\frac{R_{BLR}}{R_g}\right)^{1/2}\left(\frac{FWHM}{c}\right),
\label{eq 28}
\end{equation}
\noindent where $FWHM$ is the observed full width of the emission line and $R_g=GM_{BH}/c^2$ is the gravitational radius.

The values for the inclination angle $i$ for a number of chosen objects corresponds to those obtained by expression (28). These values are presented in Table 5.

In literature there exist various estimates of the inclination angle value. For example, for Akn~120 Zhang \& Wu (2002) gave the estimate of $i = 42^{\circ}$, but according to Patrick et al. (2012) $i = 33^{\circ +2}_{-17}$, and according to Walton et al. (2013) $i = 54^{\circ +6}_{-7}$ and $i = 47^{\circ +7}_{-6}$. Our result for Akn~120 does not contradict to Patrick et al. (2012) data. As concern to Zhang \& Wu (2002) estimate, they estimated the BLR inclination at the base of the bulge and stellar velocity dispersion, but not at the base of the BLR virial theorem itself. Walton et al. (2013) used the special interpretation of observed data, based on the X-ray reflection model consistent with the radio constraints on the inclination. This method is not the direct method for determining the inclination angle.

Polarization degrees $p(i)$ were determined in Martin et al. (1983); Berriman (1989); Smith et al. (2002). For these objects optical luminosities  $L_{disc}$ are also determined (see Cheluche 2013). For estimating $R_{BLR}$ we used the data of Greene et al. (2010). The luminosities $L_{jet}$ have been estimated with the relation $\dot M_{jet}=q_{jet}\dot M_{acc}$, obtained in Dutan \& Biermann  (2007) and in Dutan (2010). The values of accretion rates $\dot M_{acc}$ can be obtained from a known relation : $L_{bol}=\varepsilon \dot M_{acc}c^2$. The coefficient $\varepsilon$ of the radiation efficiency depends strongly on the spin of a black hole. The values for this coefficient are obtained in many works (see, for example, Novikov \& Thorne 1973; Krolik et al. 2007; Harko et al. 2009). After deriving $\dot M_{acc}$ from determined bolometric luminosities of chosen objects (Ho et al. 2008; Brenneman 2013) we determined $\dot M_{jet}$ using the data of Dutan \& Biermann  (2007) and Dutan (2010). Then we obtained optical $L_{jet}$ according to relation: $L_{jet}=\varepsilon\dot M_{jet}c^2$.

It is very important that in the magnetic connection model for launching jets from Kerr black holes the energy power of a jet does not depend strongly on the mass of the black hole, because the magnetic energy density at the horizon radius  $B_H^2\sim M_{BH}^{-2}$. For Blandford-Znajek (BZ) mechanism of jet generation  $L_{jet}(total)\sim B^2_HM^2_{BH}a^2$ does not depend on the black hole mass, especially in the case of equipartition between densities of magnetic and accreting matter energies.
We take the sources with small deviation ($\Delta PA < 24 ^{\circ}$) of position angle from the axis of radio emission ( see Table 4 in Smith at al. 2004).

Very frequently one assumes that optical radiation from an accretion disc is non-polarized. As was mentioned above, this occurs  due to Faraday depolarization effect (see also, Silant'ev et al. 2009, 2013). Besides we assume that the jet is optically thin. Under these  assumptions the degree of ''negative'' polarization can be calculated according to the formula:

\begin{equation}
 p(i)=\frac{p_{point}(i)}{1+ L_{disc}(i)/L_{jet}(i)}.
 \label{eq29}
\end{equation}

\noindent Recall, that $\cos i=\mu$ and the polarization degree $p_{point}(\mu)$  is presented in Tables 2 and 3.

The data on optical luminosities for standard accretion discs  in AGNs are presented in Chelouche (2013). The luminosity of a non-relativistic jet core can be estimated from the usual expression ($L_{jet}=\varepsilon \dot M_{jet}c^2$), using the jet energy efficiency $\varepsilon$ presented by Krolik (2007). According to Dutan (2010), the disc particles can form the jet if and only if the BH spin parameter $a>0.755$. In Fig.4 of his paper the jet mass flow parameter $q_j$ is presented as the function of spin $a$. We used these data for estimating the luminosity $L_{jet}$.

In Table 5 we present the estimates of polarization degrees $p(i)$ for a number of Seyfert-1 AGNs calculated from formula (29).  Our calculations  give practically the same values of polarizations as the observed ones:(for Akn120 $ p\simeq 0.79-0.35\%$, for Mrk1048 $p\simeq 0.8\pm 0.2$, for Mrk509 $p\simeq 0.85 - 0.35\%$, for Mrk6 $p\simeq 0.9\%$, for NGC 3516 $p\simeq 0.83 -0.72\%$,
for NGC 4051 $p\simeq 0.67 -0.55\%$, for NGC 4151 $p\simeq 0.28\%$).

\begin{table}
\caption{Polarization $p(i)$ of some Seyfert-1 AGNs.}
\begin{tabular}{|p{1.1cm}|p{0.5cm}|p{0.5cm}|p{0.4cm}|p{0.4cm}|p{0.4cm}|p{0.6cm}|p{0.4cm}|p{0.3cm}|}
\hline
$ Source $  & $\Delta PA^{\circ}$&$ i^{\circ}$  &$ Spin$ & $q_j$ & $\frac{L_{disc}}{L_{jet}}$ & $p_{point}$ & $p(i)$ & $Ref$ \\
\hline
  $Akn 120$ & 24  & 29    & 0.95   & 0.1   &  1.59   & 1.07    & 0.41 &$( 1 )$\\
  $Mrk1048$  & 1.5 & 39.5 & 0.9 & 0.1 & 1.0 & 1.91 & 0.95  & $(2)$\\
  $Mrk 509$   & 20.9 & 43 & 0.86 & 0.05 & 2.0 & 2.0 & 0.67  & $(1)$ \\
  $Mrk 6$ & 13.5 & 41.7 & 0.99 & 0.6 & 2.29 & 3.04 & 0.92  & $(1)$ \\
  $ NGC 3516$ & 17 & 39 & 0.9 & 0.1 & 1.66& 1.83 & 0.69  & $(2)$ \\
  $NGC 4051$  & 12 & 32 & 0.9 & 0.1 & 1.0 &1.27 & 0.64 & $(3)$ \\
  $NGC 4151$  & 14 & 23.5 & 0.88& 0.05& 1.86 & 0.79 & 0.28  & $(2)$ \\

\hline
\end{tabular}
\end{table}

\section{Conclusion}

The calculations in our paper demonstrate that the ''negative'' polarization of radiation from Seyfert-1 AGNs, corresponding to wave electric field oscillations in the plane $({\bf nN})$ (the line of sight ${\bf n}$ - the normal to accretion disc ${\bf N}$), can be explained not only in the model with an optically thin accretion disc ( or equatorial flow) but also when the accretion disc is optically thick. Our calculations show that the reflection of non-polarized radiation from the central jet in an optically thick accretion disc always produces ''negative'' polarization. The polarization of radiation emerging from the optically thick accretion disc  (the Milne problem) has positional angle perpendicular to plane $({\bf nN})$. The total polarization of Seyfer-1 AGN depends on the relative values of  luminosities of radiation from the jet and the accretion disc. We have shown that the observed polarization of Seyfert -1 AGN is valid even  though the luminosity of disc is considerably higher than that of the jet.  Allowing for Faraday depolarization in the accretion disc decreases the perpendicular polarization of radiation in Milne problem.  The cases with intermediate positional angles (at an angle to the plane( ${\bf nN}$))  may be explained by presence of Faraday rotations in the accretion disc. For axially symmetric model this occurs if there is a magnetic component perpendicular to the disc. We present the calculation of polarization degrees for a number of Seyfert-1 AGNs. The results of our calculations are close to the observed values of polarization.

\section*{Acknowledgments}

This research was supported by the Basic Research Program P-7 of Presidium of RAS and the program of the Department of Physical Sciences of RAS No.2.

\label{lastpage}

\end{document}